\documentclass[notitlepage,superscriptaddress,showpacs,nobalancelastpage,twocolumn,aps,longbibliography,prb]{revtex4-2}
\RequirePackage[T1]{fontenc}
\RequirePackage{times} % favourite for note.

\usepackage{siunitx} % typesets numbers with units very nicely
\usepackage[italicdiff]{physics}
\usepackage{amsfonts}
\usepackage{lipsum}
\usepackage{multirow}
\usepackage{subfigure}
\usepackage{amsmath}
\usepackage{amssymb}
\usepackage{graphicx,epstopdf}
\usepackage{xspace}
\usepackage{array}
\usepackage[hidelinks]{hyperref}
\usepackage[dvipsnames]{xcolor}
\usepackage{my_symbols}
\usepackage{CJK}
\usepackage{bm}
\usepackage{verbatim}
\usepackage{tikz}
\usepackage{comment}
\usepackage{marginnote}
\usepackage[textwidth=1.5cm]{todonotes}
\usetikzlibrary{calc,3d}
\usetikzlibrary{arrows}
\usepackage[normalem]{ulem}
\usepackage{soul}
\usepackage{dsfont}
\usepackage{accents}

\sethlcolor{green}

{\par}

\newcounter{mycomment}

\includecomment{comment} %show comments
\begin{document}

\begin{CJK*}{UTF8}{gbsn} % Use default fonts from CJK (see below)
\title{Anatomy of Spin Wave Polarization in Ferromagnets}
% \author{Yutian Wang}
\author{Yutian Wang (王昱天)}
\author{Ruoban Ma}
\affiliation{Department of Physics and State Key Laboratory of Surface Physics, Fudan University, Shanghai 200433, China}
\author{Jiang Xiao (萧江)}
\email[Corresponding author:~]{xiaojiang@fudan.edu.cn}
\affiliation{Department of Physics and State Key Laboratory of Surface Physics, Fudan University, Shanghai 200433, China}
\affiliation{Institute for Nanoelectronics Devices and Quantum Computing, Fudan University, Shanghai 200433, China}
%\affiliation{Shanghai Qi Zhi Institute, 200232 Shanghai, China}
\affiliation{Shanghai Research Center for Quantum Sciences, Shanghai 201315, China}
%\affiliation{Zhangjiang Fudan International Innovation Center, Fudan University, Shanghai 201210, China}
\affiliation{Hefei National Laboratory, Hefei 230088, China}
% \affiliation{Shanghai Branch, Hefei National Laboratory, Shanghai 201315, China}

\begin{abstract}

    Spin waves in ferromagnetic materials are predominantly characterized by right-handed circular polarization due to symmetry breaking induced by net magnetization. However, magnetic interactions, including the external magnetic field, Heisenberg exchange, Dzyaloshinskii-Moriya interaction, and dipole-dipole interaction, can modify this behavior, leading to elliptical polarization. This study provides a systematic analysis of these interactions and their influence on spin wave polarization, establishing principles to predict traits such as polarization degree and orientation based on equilibrium magnetization textures. The framework is applied to diverse magnetic configurations, including spin spirals, domain walls, and Skyrmions, offering a comprehensive yet simple approach to understanding polarization dynamics in ferromagnetic systems.

\end{abstract}
\maketitle
\end{CJK*}

\section{Introduction} 

Spin waves \cite{stancilSpinWavesTheory2009}, or magnons, represent collective excitations of magnetic moments in magnetic materials, playing an important role in many aspects of condensed matter physics, and being most critical in the emerging field of magnonics. 
\cite{patton_magnetic_1984,miyazaki_physics_2012,demokritovMagnonicsFundamentalsApplications2012, chumak_magnon_2015,yuMagneticTextureBased2021,yuanQuantumMagnonicsWhen2022,chen_magnon_2024,   zarerameshtiCavityMagnonics2022}. 
% \wyt{Does here need fundamental theoretical textbook of spin wave and magnon?}
% \jx{Yes, add Stancil's Spin Wave book: Stancil, D. D. Prabhakar, A. Spin Waves. (Springer, Office, 2009).}
% \wyt{The ref[6-9] are references of a deleted sentence. They correspond to ref[5-8] in V0.3. Should we keep them?}
Their distinctive attributes - low dissipation, a broad frequency range spanning from GHz to THz, and the capability to propagate through magnetic conductors, semiconductors, and insulators - position spin waves as a promising candidate for the development of future energy-efficient information technologies \cite{barman2021MagnonicsRoadmap2021}.

% Spin waves, or magnon, the collective excitations of magnetic moments in magnetic materials, are essential for understanding various aspects of condensed matter physics 
% especially in the emerging field of magnonics  
% The unique combination of properties such as the low dissipation, wide frequency range from GHz to THz, and the ability to propagate in magnetic conductor, semiconductor, and insulator, make spin waves a promising candidate for the development of future energy-efficient information technologies \cite{barman2021MagnonicsRoadmap2021}. 
% The unique ability of spin waves to propagate through magnetic conductors, semiconductors, and even insulators positions them as a promising candidate for the development of future energy-efficient information technologies \cite{barman2021MagnonicsRoadmap2021}.

Similar to acoustic and optical waves, the spin wave also has the polarization degree of freedom, which reflects the oscillatory direction in the dynamics of magnetic moments. In ferromagnetic materials, spin waves predominantly exhibit right-handed circular or elliptical polarization \cite{herringTheorySpinWaves1951}, a consequence of symmetry breaking due to net magnetization. Conversely, antiferromagnetic spin waves can display both left-handed and right-handed polarizations \cite{kefferTheoryAntiferromagneticResonance1952}, as well as various combinations, leading to all kinds of linearly or elliptically polarized states. This versatility in antiferromagnetic spin wave polarization has spurred numerous theoretical and experimental investigations aimed at manipulating these polarizations, including applications such as transmitting spins via polarized spin waves \cite{hanBirefringencelikeSpinTransport2020}, implementing magnetic logic gates \cite{yuMagneticLogicGate2020}, and developing spin wave polarizers \cite{lanAntiferromagneticDomainWall2017} and spin wave field-effect transistors \cite{chengAntiferromagneticSpinWave2016}. In contrast, the manipulation of spin wave polarization in ferromagnetic materials is comparatively limited, as it is largely constrained to right-handed polarization by the direction of net magnetization. 
Nevertheless, there remain significant opportunities for exploration of the polarization properties of ferromagnetic spin waves, particularly in the context of efficient magnon injection and the interaction between ferromagnetic spin waves and microwave photons \cite{zarerameshtiCavityMagnonics2022,zhang_strongly_2014,bhoi_photon-magnon_2019} or phonons \cite{berk_strongly_2019,hioki_coherent_2022}. There are studies suggest that the polarization of ferromagnetic spin waves are linked to their dissipation \cite{patton_linewidth_1968,kambersky_spin-wave_1975,kalarickalFerromagneticResonanceLinewidth2006,rozsaEffectiveDampingEnhancement2018}, efficiency of spin pumping \cite{andoOptimumConditionSpincurrent2009,nonakaSpinPumpingMagnetizationprecession2011} and the angular momentum carried by magnon \cite{kamraNonintegerspinMagnonicExcitations2017,bauerSoftMagnonsAnisotropic2023}, underscoring the importance of further understanding ferromagnetic spin wave polarization.
% \jx{Most of the references are antiferromagnetic spin wave, are there any ferromagnetic spin wave examples?}
% \jx{Cite Gerrit's soft magnon paper somewhere.}
% \wyt{Ok.}

The spin wave polarization characteristics in ferromagnets are generally limited, predominantly exhibiting right-handed circular polarization in simplified scenarios. However, various interaction mechanisms can cause deviations, leading to elliptically polarized spin waves that maintain their overall right-handed orientation. While it is well known that magnetocrystalline anisotropies can induce non-circular polarization when the anisotropy axis misaligns with the equilibrium magnetization \cite{hillebrands_spin_2003}, it is less recognized that spin wave excitations in inhomogeneous magnetic textures also tend to exhibit elliptically polarized behavior \cite{kravchukSpinEigenmodesMagnetic2018,rozsaEffectiveDampingEnhancement2018}. Despite these insights, the current understanding of non-circular polarization is largely incomplete, lacking a comprehensive and systematic framework to predict its occurrence and more importantly its specific orientation of the elliptical polarization. The complexity increases with multiple magnetic interactions and intricate magnetic textures. It would be very helpful to have general principles that can reliably determine spin wave polarization in ferromagnet, and more importantly distinguishing the different roles of various interactions on spin wave polarization.

In this paper, we present a systematic approach to analyzing ferromagnetic spin wave polarization by breaking down the complexities of various magnetic interactions into fundamental components. This decomposition allows us to formulate a set of spin wave polarization rules that can be universally applied to understand more intricate scenarios. When multiple magnetic interactions are involved, the overall polarization can be effectively deduced by aggregating the individual contributions of each interaction. 
%By establishing this framework, we not only provide a clearer insight into the polarization behavior of spin waves in ferromagnetic systems, \add{but also a straightforward way to qualitatively determine the spin wave polarization without the reliance on complex numerical simulations. By streamlining both analysis and device design, it offers researchers a more efficient pathway for theoretical investigation and experimental implementation.}
This work offers a simplified approach to understanding and determining spin wave polarization in ferromagnetic systems, circumventing the need for complex numerical simulations. By providing a clearer insight into polarization behavior, it streamlines both theoretical analysis and device design, enabling more efficient research and experimental implementation for researchers in the field.

This paper is organized as the following: In Section II, we employ free energy analysis as a foundational tool to examine the influences of each type of magnetic interaction on the spin wave polarization. Section III distills the spin wave polarization behavior into a concise set of guiding rules. Section IV presents several illustrative examples that demonstrate the application of these rules. Finally, we conclude with a summary that encapsulates the key findings and implications.

\section{Spin Wave Polarization influenced by various interactions}

The magnetization dynamics in ferromagnets is described by the Landau-Lifshitz-Gilbert equation \cite{landau_theory_1992, gilbert_classics_2004},
\begin{equation}
    \label{eqn:LLG}
    \dot \mb(\br,t) = - \gamma \mb \times \bH_{\rm eff}+\alpha \mb\times\dot\mb,
\end{equation} 
where $\gamma$ and $\alpha$ are the gyromagnetic ratio and the Gilbert damping parameter. In linear response regime, the local magnetization undergoes a precessional motion around its equilibrium direction determined by the effective magnetic field $\bH_{\rm eff} = -\delta F/\delta \mb$, where $F$ is the free energy of the system. The precessional motion traces out a trajectory in the plane perpendicular to the equilibrium magnetization. The shape of the trajectory characterizes the polarization of the spin wave excitation, which can be linearly, circularly, or elliptically polarized. The precession direction with respect to the equilibrium magnetization defines the handedness of the polarization, which can be right-handed or left-handed. 
% Just like in the phonon polarization or optical polarization, the spin wave polarization relies on the trajectory of the precessing magnetic moment, which in most cases are elliptically shaped. 
% The spin wave polarization can be quantified using the ellipticity of the spin wave trajectory:
% \begin{equation}
%     \label{eqn:eta}
%     \eta = \frac{b}{a}
% %    r \equiv \half \log\frac{a}{b}.
% \end{equation}
% where $a$ and $b$ are the semi-axis along the two principal axes of the elliptical trajectory.  
% Evidently, $\eta = \pm 1$ corresponds to the left and right circular polarization, and $\eta = 0, \infty$ correspond to the linear polarization along the two principal axes.

% \begin{figure*}[t]
%     \includegraphics[width=\textwidth]{models.png}
%     \caption{Dispersions and linewidth for (a) and (b) 1D (anti)ferromagnet, (c) and (d) 1D ferromagentic ring, (e) and (f) 1D antiferromagnetic ring, (g) and (h) 2D magnetic Skyrmion.}
% \end{figure*}

In this paper, we focus exclusively on the linearized spin wave regime, which allows us to treat the effects of various magnetic interactions - namely, the external magnetic field, magnetic anisotropies, Heisenberg exchange interaction, Dzyaloshinskii-Moriya interaction, and dipole-dipole interactions - as independent and additive factors. This approach facilitates a detailed analysis of each interaction's contribution to spin wave excitation and polarization. 
Not only the influence of these interactions on polarization depend on the inherent nature of each interaction, but also on the specific configuration of the magnetic ground state, highlighting the intricate interplay between these elements in spin wave phenomena.

To simplify the analysis, we only keep the absolutely minimal model with three considerations: First, we assume the long wave length approximation. The situation with (not-so-extreme) finite wave vector shall only have quantitative differences. Second, upon a given equilibrium magnetization distribution $\mb_0(\br)$ stabilized by considering all interactions, we shall single out one interaction and analyze the effect of this interaction on the polarization. Last, we assume a 'virtual' excitation of circular polarization upon $\mb_0(\br)$, and work out the energy as function of the excitation direction, and the direction with lower energy represents the polarization elongation direction. With the last consideration, we let the local magnetization slightly deviates from its equilibrium magnetization $\mb_0$ (assumed to be $\hbz$) direction by an angle $0 < \theta \ll 1$ in the direction of $\phi$:
\begin{equation} 
    \mb = (\sin\theta\cos\phi, \sin\theta\sin\phi, \cos\theta) \equiv \bn_{\theta,\phi}.
\end{equation}
The free energy change for the interaction singled out due to this virtual deviation can be expanded in terms of the angular deviation $\theta, \phi$ as: 
\begin{equation}
    \label{eqn:F}
    F = f(\phi) \theta^2 + O(1,\theta^3, \theta^4 \dots).
\end{equation}
The $O(1)$ term represents an overall energy shift and can be ignored. The linear therm $O(\theta)$ shall be balanced by other interactions because $\hbz$ (or $\theta = 0$) has been assumed to be the equilibrium direction. The $O(\theta^3)$ and higher order terms are neglected because we only consider small deviations $\theta \ll 1$. 
Consequently, the spin wave polarization is determined by the angular dependence of the quadratic coefficient $f(\phi)$, the minimums of which denote the polarization elongation direction. We should note that \Eq{eqn:F} is the excitation energy on a pre-existing magnetic texture, while the free energy that is responsible for the formation of the underlying texture is not included. 

In all cases discussed below, the coefficient of the $\theta^2$ term $f(\phi)$ in \Eq{eqn:F} can be decomposed into the $\phi$-independent and -dependent parts in the following form
\begin{equation}
    \label{eqn:f}
    f(\phi) = \epsilon - \abs{s} \cos 2(\phi_\text{m}-\phi),
\end{equation}
where $\epsilon$ is the coefficient of $\theta^2$ that is $\phi$-independent, and the cosine term has the capability in expressing any $\phi$ dependent functions that is quadratic in $\sin\phi, \cos\phi$. This form means that the free energy $F$ in \Eq{eqn:F} is squeezed in the $\phi = \phi_m$ direction by the strength $\abs{s}$, \ie the quadratic curvatures are $\epsilon\pm\abs{s}$ along the major and minor axes in $\phi = \phi_m$ and $\phi = \phi_m + \pi/2$, respectively. Thereby, the spin wave excitation shall be elongated in the $\phi = \phi_m$ direction. We define the squeezing parameter as
% \jx{Is this classical squeezing parameter defined exactly the same as the one defined in quantum one?}
% \wyt{Yes. Up to a phase factor at most.}
\begin{equation}
    \label{eqn:s}
    s \equiv -\abs{s} e^{2i\phi_\text{m}},
\end{equation}
whose phase and strength determine the ellipticity of the spin wave polarization and its orientation, respectively.

The resulting elliptical trajectory is
\begin{equation}
    \theta(\phi) \propto \frac{1}{\sqrt{f(\phi)}}
    = \frac{1}{\sqrt{\epsilon + \Re{s e^{-2i\phi}}}},
\end{equation}
which has ellipticity 
\begin{equation}
    \label{eqn:eta}
    \zeta = \frac{a^2-b^2}{a^2+b^2} = \frac{\abs{s}}{\epsilon},
    % = \sqrt{\frac{\epsilon+\abs{s}}{\epsilon-\abs{s}}}
    % = \frac{\epsilon+\abs{s}}{\omega} 
    % = \sqrt{1+\abs{\frac{s}{\omega}}^2}+\abs{\frac{s}{\omega}}.
\end{equation}
and the major axis is in the $\phi_m$ direction. Here $a, b$ are the lengths of the semi-major and minor axis of the ellipse.
%where $\omega = \sqrt{\epsilon^2 - s^2}$ is the eigen freuquency with free energy given by \Eq{eqn:F} and \Eq{eqn:f}.

% is a linear combination of quadratic in $\sin\phi, \cos\phi$, thus the resulting equal-energy trajectory shall correspond to a ellipse trajectory for $\theta(\phi)$ with major and minor axis direction is given by the minimum and maximum value of $f(\phi)$.
% The final polarization shall be concluded by considering the effects from all interactions. 
% To characterize the ellipticity of the polarization, we define the modulation parameter $s$ for $f(\phi)$:
% \begin{equation}
%     s = \qty[f(\phi_\text{max}) - f(\phi_\text{min})]e^{i2\phi_\text{max}},
% \end{equation}
% where $\phi_\text{max}$ and $\phi_\text{min}$ are the angles corresponding to the maximum and minimum value of $f(\phi)$, respectively.

\begin{figure*}[t]
    \includegraphics[width=\textwidth]{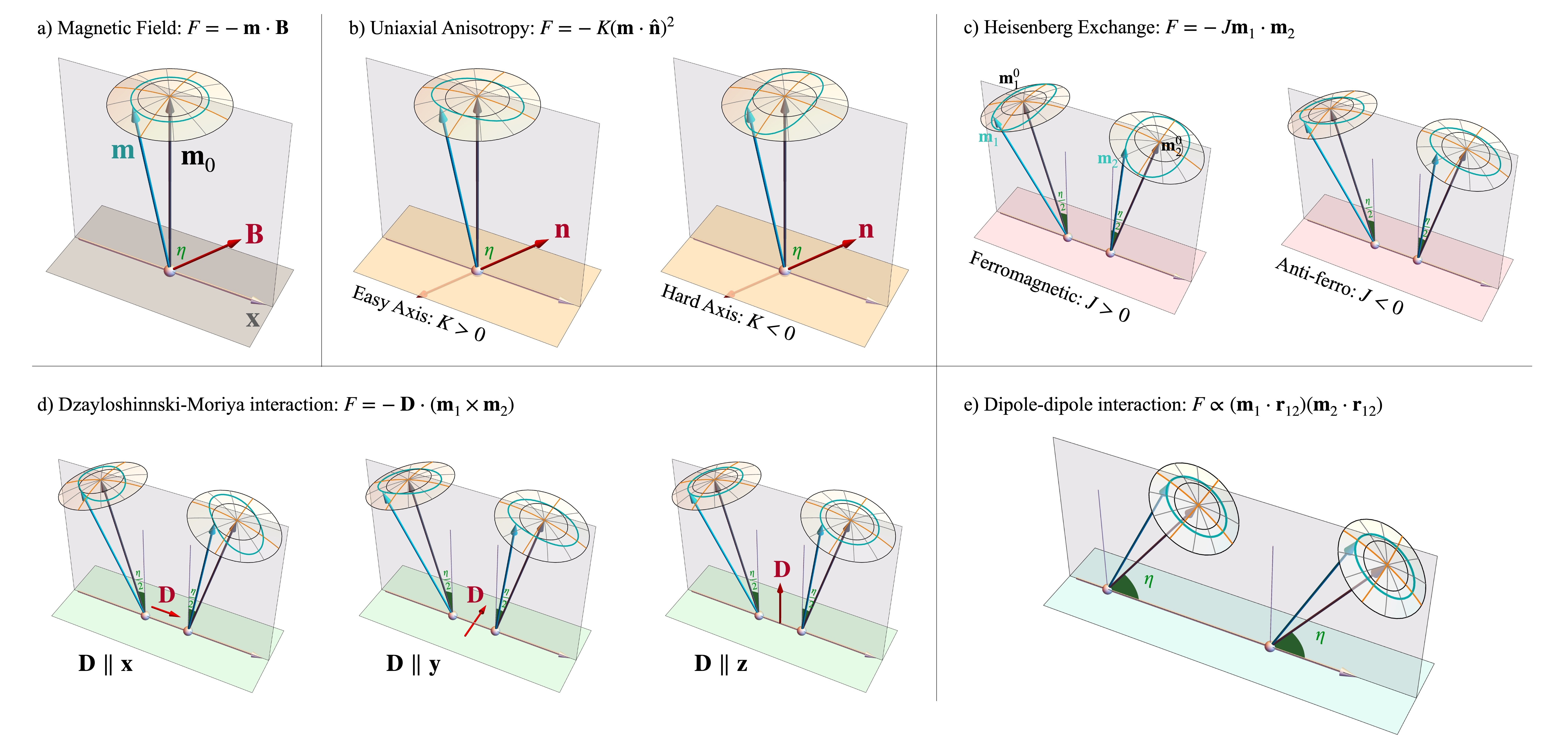}
    \caption{Influence of various interactions on the spin wave polarization.}
    \label{fig:polarization}
\end{figure*}

\emph{Magnetic field - } With the general considerations stated above, we shall first work out the simplest case: how would an external magnetic field influence a presumed circularly polarization spin wave in a macrospin model. As depicted in \Figure{fig:polarization}(a), we assume the equilibrium magnetization points in $\mb_0 = \hbz$ stabilized by all interactions, among which the external magnetic field $\bB = B\bn_{\eta,0} = B(\sin\eta, 0 , \cos\eta)$ is applied with an angle $\eta$ away from $\hbz$. The free energy due to the applied magnetic field alone is $F_B = - M \mb \cdot \bB$. Let the magnetization having a virtual deviation from the equilibrium by a small angle $\theta$ in the direction of $\phi$, \ie $\mb = \bn_{\theta,\phi}$. We find $F_B =  -MB \cos\eta~\theta^2/2$ and $f(\phi) = -MB\cos\eta/2$ is independent of $\phi$, therefore the corresponding squeezing parameter for the applied field vanishes: $s_\text{B} = 0$. Therefore, the excitation upon $\mb_0$ keeps its original (right-handed) circular polarization, as depicted in \Figure{fig:polarization}(a). We then conclude that, the external field, regardless of its direction relative to the equilibrium magnetization direction, does not affect the spin wave polarization. In other words, the spin wave polarization, if deviates from circular, should not be attributed to the magnetic field, but to other interactions.
% \wyt{We need another letter to represent the angle instead of $\gamma$, maybe $\xi$?}
% \jx{The last sentence is added because I am afraid that readers may think that the external field can influence the polarization when applied non-collinear to the magnetization. But this is a misunderstanding because such non-collinear field is to first modify the equilibrium magnetization, then other interactions come in play polarization the spin wave. Basically, the scenarios in b). But readers may not grasp this point at the first glance.}
% \jx{Because $\alpha$ is used for Gilbert damping, I changed all angle $\alpha$ to angle $\gamma$. Even though it is still conficts with the gyromagnetic ratio, but I guess this would be less confusion.}
% \wyt{OK.}

\emph{Uniaxial Anisotropy - } We now consider the effect of uniaxial anisotropy on the spin wave polarization. We assume the anisotropy axis points along $\bn = (\sin\eta, 0, \cos\eta)$, and the associated anisotropy energy is $F_K = - K (\mb\cdot\bn)^2/2$. The axis is an easy axis for $K > 0$ and a hard axis for $K < 0$. Again, we assume the equilibrium magnetization points along $\mb_0 = \hbz$, and let the magnetization have a virtual deviation from the equilibrium by a small angle $\theta$ in the direction of $\phi$ with $\mb = \bn_{\theta,\phi}$, the resulting $\phi$-depedent anisotropy energy that is quadratic in the deviation angle $\theta$ is
\footnote{The $\phi$-independent term in the anisotropy energy is ignored.}
\begin{equation}
    \label{eqn:sK}
    % F_K = -K\qty(\sin^2\eta\cos^2\phi - \cos^2\eta) \theta^2 
    F_K = -\frac{K}{2}\sin^2\eta\cos^2\phi~\theta^2 
    \qRa s_\text{K} = -\frac{K}{2}\sin^2\eta
\end{equation}
% for which $s_\text{K} = -K\sin^2\eta$. 
% \wyt{By define, $s$ should be non-negative}
% \jx{OK, but how to keep the sign meaningful? Maybe we can define the $s$ in such a way that it is positive for polarizing in the texture plane or $\mb$-$\bn$ plane. }
% \wyt{Adding an extra phase will help it.}
When $K >0$, \ie the easy axis case, according to \Eq{eqn:s}, we have $\phi_m = 0$ and the energy minimizes at $\phi = 0, \pi$. Therefore, the spin wave tends to elongate in the plane spanned by the anisotropy axis $\hbn$ and the equilibrium magnetization $\mb_0$ as seen in the left panel of \Figure{fig:polarization}(b). On the other hand, when $K < 0$, \ie $\hbn$ is a hard axis, we have $\phi_m = \pi/2$ and the energy minimizes at $\phi = \pi/2, 3\pi/2$, thus the spin wave tends to elongate in the direction perpendicular to the plane as shown in the right panel of \Figure{fig:polarization}(b). 

\emph{Heisenberg exchange - } A slightly more complicated situation is to the effect of the exchange interaction on the spin wave polarization. We shall see that such influence only happens when there is inhomogeneous (or non-collinear) magnetic texture. To its theoretical minimum, it is sufficient to consider the Heisenberg exchange between two lattice points, for which the exchange free energy $F_H = - J \mb_1\cdot\mb_2$. 
We assume that the equilibrium magnetization from the two lattice sites form an angle $\eta$: $\angle(\mb_1^0, \mb_2^0) = \eta$. We define the average direction of the equilibrium magnetization from the two lattice sites as $\hbz \propto \mb_1^0 + \mb_2^0$, and the plane spanned by $\mb_1^0$ and $\mb_2^0$ as $x$-$z$ plane, or termed as the texture plane. Then the two equilibrium magnetizations tilt away from $\hbz$ by angle $\pm\eta/2$ in the texture plane, and can be expressed as:
% 1) $\hz$ the direction of $\mb_1+\mb_2$, 2) $\hx$ the direction of magnetization gradient, \ie $\hx = \frac{\mb_2-\mb_1}{|\mb_2-\mb_1|}$, 3) $\hy$ satisfies $\hy=\hx \times \hz$. The equilibrium is then in the $\hx-\hz$ plane, tilting away from $\hz$ by a small angle $\pm\frac{\eta}{2}$:
% \wyt{I think it will be natural to construct the local coordinate from the magnetization given. $\eta$ is always positive under this definition.}
% \jx{I prefer to define the coordinates in a more natural way with less strict specification.}
% We assume the equilibrium magnetization for both are in the $\hbx-\hbz$ plane, tilting away from $\hbz$ by a small angle $\pm\frac{\eta}{2}$: 
% \jx{Maybe $\pm/\eta/2$ is more natural, update for this change.}
$\mb_1^0 = \bn_{-\eta/2,0}$ and $\mb_2^0 = \bn_{\eta/2,0}$. When $\eta = 0$, \ie the equilibrium magnetization is uniform, the Heisenberg exchange energy due to homogeneous excitation $\mb_1 = \mb_2 = \bn_{\theta,\phi}$ is constant $F_H = - J$ and has no preference in the excitation direction. The effect of Heisenberg exchange on the spin wave polarization only appears when the equilibrium magnetization is non-collinear, \ie $\eta \neq 0$. 
The excitation upon \emph{inhomogeneous} background is achieved by rotating the excited magnetization ($\bn_{\theta,\phi}$) upon \emph{uniform} background along $\hby$ by angle $-\frac{\eta}{2}$ and $+\frac{\eta}{2}$, \ie $\mb_1 = R_{-\eta/2}^y \bn_{\theta,\phi}$ and $\mb_2 = R_{\eta/2}^y \bn_{\theta,\phi}$, where $R_\chi^y$ is a rotation matrix corresponding to rotate angle $\chi$ about $\hby$ direction. Then the Heisenberg exchange energy due to this virtual excitation is 
% \jx{I've checked that by including the wave vector dependence $q$ (by inserting $R_{\pm qa}^z$ before $\bn_{\theta,\phi}$), there won't be any term that depends on both $q$ and $\phi$, therefore the conclusions doesn't seem to change. }
\begin{align}
    \label{eqn:sH}
    F_H &= - J (R_{-\eta/2}^y \bn_{\theta,\phi})\cdot
    (R_{\eta/2}^y \bn_{\theta,\phi}) \nn
    &\simeq -2J\sin^2\frac{\eta}{2}\sin^2\phi~\theta^2
    \qRa s_\text{H} = 2J\sin^2\frac{\eta}{2}.
\end{align}
For ferromagnetic coupling ($J > 0$), this energy minimizes at $\phi = \phi_m = \pm\pi/2$, therefore the ferromagnetic Heisenberg exchange tends to squeeze the spin wave to be polarized perpendicular to texture plane, as shown in left panel of \Figure{fig:polarization}(c). For antiferromagnetic coupling ($J < 0$), the presumed soft texture (with small $\eta$) tends to confine the spin wave within the texture plane, as shown in right panel of \Figure{fig:polarization}(c). 

Here, we assumed the same deviation angle $\theta$, $\phi$ to parametrize the magnetic deviations for both $\mb_1$ and $\mb_2$ from their respective (non-collinear) equilibrium directions. Such simplification is to effectively consider the macrospin-like excitation in long wavelength approximation. Such virtual macrospin-like excitation in non-collinear magnetic background is achieved by rotating a real macrospin excitation up on a uniform background in such a way that the non-collinear background is reproduced by the rotation. In general, there is a difference between the two kinds of excitations: rotating the background first then exciting versus exciting first then rotating. However, their difference is negligible when the excitation amplitude and the texture gradient are both small, which is the situation we consider here: $\theta, \eta \ll 1$.

\emph{Dzyaloshinskii-Moriya interaction (DMI) - } We now consider the more complex case of asymmetric exchange coupling, \ie Dzyaloshinskii-Moriya interaction defined by the interaction free energy $F_\text{DM} = -\bD\cdot(\mb_1\times\mb_2)$.
%\add{with $\bD = \bD_\rho + D_z\hz=(D_\rho\cos\phi_D,D_\rho\sin\phi_D,D_z)$}.
%We shall assume here that the equilibrium magnetization texture is favored by the DMI, \ie $\bD\cdot(\mb_1^0\times\mb_2^0) \le 0$.
Based on the same procedure as the Heisenberg exchange, the energy associated with the finite magnetic texture ($\mb_1^0 \neq \mb_2^0$) is expressed as
\begin{align}
    F_\text{DM} &= -\bD \cdot \qty[(R_{-\eta/2}^y \bn_{\theta,\phi})
    \times (R_{\eta/2}^y \bn_{\theta,\phi})] \nn
    &\simeq \qty[\bD \cdot (\sin\frac{\eta}{2}\sin 2\phi,\sin \eta\sin^2\phi,0)] \theta^2,
\end{align}
from which we may find that the corresponding squeezing parameter relies on the direction of the DM vector $\bD$:
\begin{equation}
    \label{eqn:sD}
    s_\text{DM} = 
    \begin{cases}
    -iD\sin(\eta/2) & \bD \parallel \hbx, \\
    -D\sin\eta & \bD \parallel \hby, \\
    0 & \bD \parallel \hbz. 
    \end{cases}
\end{equation}
When $\bD = D\hbx$ with $D>0$, \ie parallel to the gradient direction of $\mb_i^0$ within the texture plane, the spin wave tends to elongate in the direction $\phi_m = -\pi/4$ (when $\eta > 0$), as shown in the left panel of \Figure{fig:polarization}(d). 
% \jx{What about $D<0$?}
% \wyt{The polarization plane will be rotated by $\pi/2$ as long as the ground state is the same.}
When $\bD = D\hby$, \ie perpendicular to the texture plane, and the equilibrium magnetic texture is consistent with the DMI ($-\bD\cdot(\mb_1^0\times\mb_2^0) \le 0$ or $\eta \ge 0$), the spin wave is elongated in the direction $\phi_m = 0$, \ie within the texture plane, as shown in middle panel of \Figure{fig:polarization}(d). Finally, when $\bD = D \hbz$, \ie in the angular bisector direction of $\mb_1^0$ and $\mb_2^0$ within the texture plane, there is no preference in $\phi$, meaning the polarization remains circular, as shown in the right panel of \Figure{fig:polarization}(d).

\emph{Dipole-dipole interaction - } The analysis above also applies to the non-local interactions such as the dipole-dipole interaction, which also influences the spin wave polarization. For simplicity, we consider two magnetic moments at equilibrium pointing in the same direction but form an angle $\eta$ relative to the displacement vector between the two, then the dipole-dipole interaction $F_{dd} \propto \mb_1\cdot\mb_2 - 3(\mb_1\cdot\hbr_{12})(\mb_2\cdot\hbr_{12})$. Based on the same procedure as above, we assume that both magnetic moments deviate from the equilibrium direction by a small angle $\theta$ in the direction $\phi$ (relative to the plane defined by $\mb_{1,2}^0$ and $\hbr_{12}$), and the resulting energy that is $\phi$-dependent and quadratic in $\theta$ is 
\begin{equation}
    \label{eqn:sdd}
    F_\text{dd} \propto -\cos^2\phi\sin^2\eta~\theta^2
    \qRa s_\text{dd} \propto -\sin^2\eta.
\end{equation}
This energy is independent of $\phi$ when the equilibrium magnetic moments are collinear with their displacement vector ($\mb_{1,2}^0 \parallel \hbr$ or $\eta = 0, \pi$). For the non-collinear case ($\mb_{1,2}^0 \not\parallel \hbr$ or $\eta \neq 0, \pi$), the excitation is elongated in the plane spanned by the equilibrium magnetic moments ($\mb_1^0 = \mb_2^0$) and the displacement vector $\hbr_{12}$, as shown in \Figure{fig:polarization}(e). And the polarization is maximal when $\eta = \pi/2, 3\pi/2$, or the equilibrium magnetization are perpendicular to the displacement vector.

\section{Spin Wave Polarization Rules}

The analysis illustrated above in \Figure{fig:polarization} can be summarized as several rules: 
\begin{itemize}
    \item Rule \#1: The uniaxial anisotropy tends to polarize the spin wave within (perpendicular to) the plane spanned by the easy (hard) anisotropy axis and the local equilibrium magnetization.

    \item Rule \#2: The Heisenberg exchange tends to polarize the spin wave perpendicular to (within) the plane spanned by magnetic texture due to (anti-)ferromagnetic Heisenberg exchange.

    \item Rule \#3: For DMI, depending on the relative relation between $\bD$, the texture plane, and the magnetization gradient within the texture plane, 
    
    \begin{itemize}
        \item Rule \#3.a: If the $\bD$ is perpendicular to the texture plane, the spin wave is polarized within (perpendicular to) the texture plane when the texture rotation is (in)consistent with $\bD$.  

        \item Rule \#3.b: If the $\bD$ is within to the texture plane and perpendicular to the magnetization gradient, then there is no effect on the polarization.
        
        \item Rule \#3.c: If the $\bD$ in the same direction as the magnetization gradient direction, the spin wave is polarized $\pm 45^\circ$ tilting from the gradient direction, and the $\pm$ sign is determined by the dot product of $\bD$ vector and the gradient direction. 
    \end{itemize}

    \item Rule \#4: For the dipolar interaction in uniformly magnetized system, the spin wave tends to polarize within the plane spanned by the magnetic moments and the displacement vector connecting the two sites.
\end{itemize}

In this study, we decompose the energy minimization for complex magnetic system into a minimization for only a pair of magnetic moments. This approach is justified, as interactions such as Heisenberg exchange, Dzyaloshinskii-Moriya (DM), and dipolar coupling are inherently pairwise and allow the total energy to be represented as a sum over spin pairs. While the application of the results for two-spin element to many-spin systems is conceptually direct-by stitching multiple pairs-this process introduces minor inaccuracies. Specifically, the optimal configuration for one pair of spins may not always align with the preferences of neighboring pairs, particularly in inhomogeneous magnetic textures. Such mismatches lead to small higher-order corrections arising from the spatially-slowly variations in magnetization; nevertheless, compared to the dominant effects of pairwise interactions, the impact on overall spin wave polarization due to the coordination of two different pairs is much smaller.
% \begin{center}
%   \begin{table*}[ht]
%   \label{tab:rules}
%   \begin{tabular}{p{3cm}||p{3cm}|p{3cm}||p{3cm}|p{3cm}} 
%   Mechanism & \multicolumn{2}{c||}{Anisotropy} & \multicolumn{2}{c}{Texture} \\ \hline
%   & easy-axis & hard-axis & Exchange & DMI \\ \hline 
%   reference plane & easy-axis + $\mb_0$ & hard-axis + $\mb_0$ & \multicolumn{2}{c}{Texture plane} \\ \hline 
%   elongation direction & in-plane & perpendicular & perpendicular & in-plane 
%   \end{tabular}
%   \caption{Rules for spin wave polarization.}
%  \end{table*}
% \end{center}

\section{Verification and Application}

Having dissected the spin wave polarization into the various influences of distinct interactions, we can now apply these established principles to more concrete cases. These include the simple macrospin model subjected to uniaxial anisotropy and an external magnetic field, the one-dimensional configuration of a textured ferromagnetic spin spiral and magnetic domain wall, as well as the two-dimensional magnetic texture of a Skyrmion. 
% Our findings indicate that the summarized rules demonstrate semi-qualitative effectiveness across these diverse systems.

\begin{figure*}[t]
    \includegraphics[width=\textwidth]{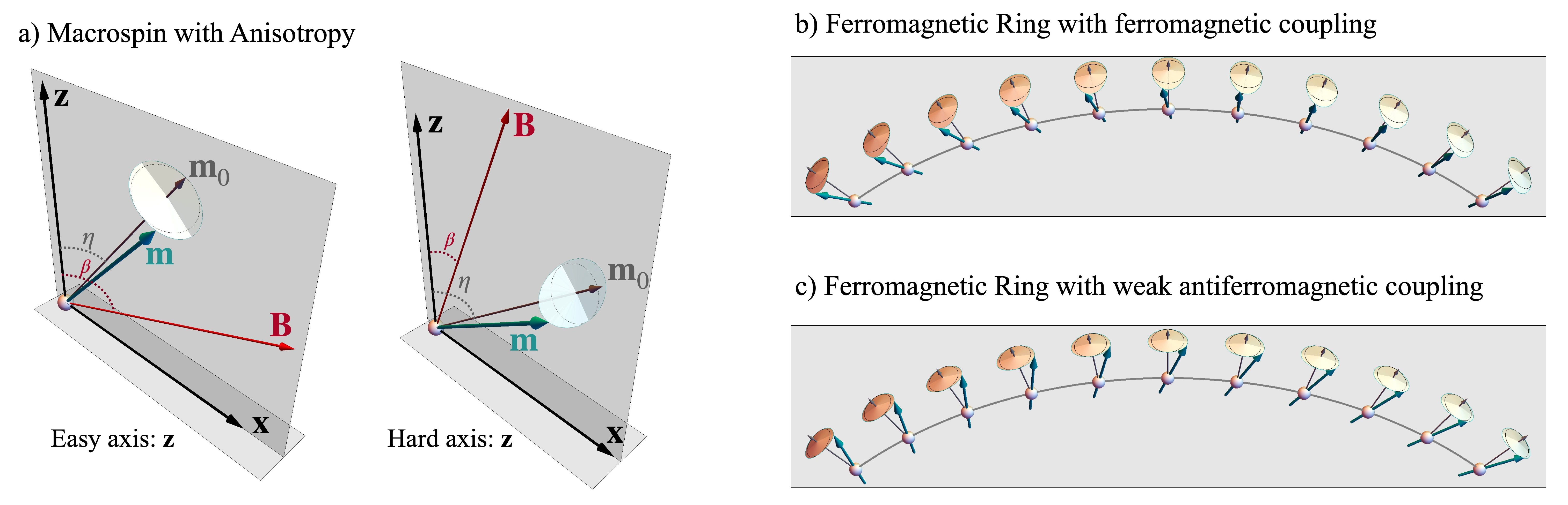}
    \caption{Depiction of spin wave polarization from the numerical simulation data. a) Macrospin excitation with $\hbz$ as the easy axis (left) and hard axis (right). b) Spin wave excitation up on a ferromagnetic ring. The major axis of spin wave polarization ellipse is perpendicular to the texture plane. c) Spin wave excitation up on a ferromagnetic ring but with weak antiferromagnetic coupling. The major axis lies within the texture plane.}
    \label{fig:sim1}
\end{figure*}

\emph{Magnetic Anisotropy - }
We consider a macrospin model with uniaxial anisotropy and a (non-collinear) external magnetic field, whose free energy is given by  
% \jx{Please check if the easy and hard axis anisotropy cases can be combined into one uniaxial anisotropy case, with $K$ being positive or negative?}
\begin{equation}
\label{eqn:F_easy}
    F = -\frac{K}{2}(\mb\cdot \hbz)^2 - \bB\cdot\mb 
\end{equation}
where $K$ is the uniaxial anisotropy along $\hbz$, and $\bB = B \bn_{\beta,0}$ tilts from the easy axis $\hbz$ by angle $\beta$. 
The equilibrium magnetization points in $\mb_0 = \bn_{\eta,0}$ with angle $\eta$ determined by $K \sin \eta \cos\eta + B\sin(\eta-\beta) = 0$ and $K\cos 2\eta + B\cos(\beta-\eta) > 0$. Since the squeezing factor from the external magnetic field vanishes $s_B = 0$, the only squeezing effect is due to the misalignment (of angle $\eta$) between the easy/hard axis and the equilibrium magnetization, which has a squeezing parameter $s_K = -(K/2) \sin^2\eta$ as given by \Eq{eqn:sK}.

\Figure{fig:sim1}(a) shows the simulated trajectory of the tip of the magnetization on a Bloch sphere with the external magnetic field ($B/\abs{K} = 1$) pointing in the red arrow direction. 
% \wyt{I want to remove the specific angle since the angle is different in hard and easy axis cases in the figure.}
% \jx{OK, do it.}
% \wyt{Is the B in the $45^\circ$ direction?}
% \jx{We can determine $\beta$ and $\eta$ when we have the real data.}
% \jx{\Figure{fig:sim1}(a) is not from real simualtion, and needs to be redraw with real simulation data.}
For $\hbz$ being the easy axis, the equilibrium magnetization lies between $\hbz$ and the $\bB$, and the resulting spin wave trajectory is elongated within the plane spanned by the easy axis and the external field. For $\hbz$ being the hard axis, the situation is the opposite, the elongation direction is perpendicular to that plane. Both behaviors are expected from Rule \#1 or \Figure{fig:polarization}(b). The simulated is carried out by solving the full LLG equation for the macrospin model numerically.
% \jx{Put label $\beta$ and $\eta$ in Figure 2a.}

\emph{Heisenberg Exchange - }
To illustrate the influence of the Heisenberg exchange on spin wave polarization, we examine a ferromagnetic ring model 
described by the following magnetic free energy
\begin{equation}
\label{eqn:F_ring}
    F = -\sum_n \qty[\frac{K}{2}(\mb_n\cdot \hbr_n)^2 
    + J\mb_n\cdot\mb_{n+1}],
\end{equation}
where $K$ is the strength of the easy-axis anisotropy along the local radial direction $\hbr_n$ of a ring with radius $R$, and $J$ denotes the Heisenberg exchange coupling between the neighboring moments along the ring. 
When $J > 0$ is ferromagnetic type and not so large ($J < K$), the magnetic ground state points to the radial direction $\mb_n^0=\hbr_n$, and the equilibrium magnetization moments of neighboring sites form an angle $\eta$: $\mb_n^0\cdot\mb_{n+1}^0 = \cos\eta$. 
Under these conditions, the easy-axis anisotropy does not produce any polarizing effect because the equilibrium magnetization consistently aligns with the local anisotropy axis throughout the system, therefore $s_\text{K} = 0$ according to Rule \#1. Thus, the only squeezing effect comes from the Heisenberg exchange, and has squeezing parameter $s_\text{H} = 2J\sin^2\eta/2$ as given by \Eq{eqn:sH}. \Figure{fig:sim1}(b) shows that the simulated spin wave excitation is polarized with its major axis perpendicular to the plane defined by the texture, as expected from Rule \#2 or the left panel of \Figure{fig:polarization}(c). 

If the Heisenberg exchange is weakly antiferromagnetic ($J < 0$ and $\abs{J} \ll K$), the equilibrium profile with $\mb_n^0 = \hbr_n$ is still quasi-stable. For such a situation, the spin wave excitation is polarized within the texture plane, as expected from Rule \#2 and the right panel of \Figure{fig:polarization}(c).
% \jx{Also add a figure for the antiferromagnetic case.}
% \wyt{I add a figure for the antiferromagnetic case. I also replace the data in figures by the real simulation data.} 

\begin{figure*}[ht]
    \includegraphics[width=\textwidth]{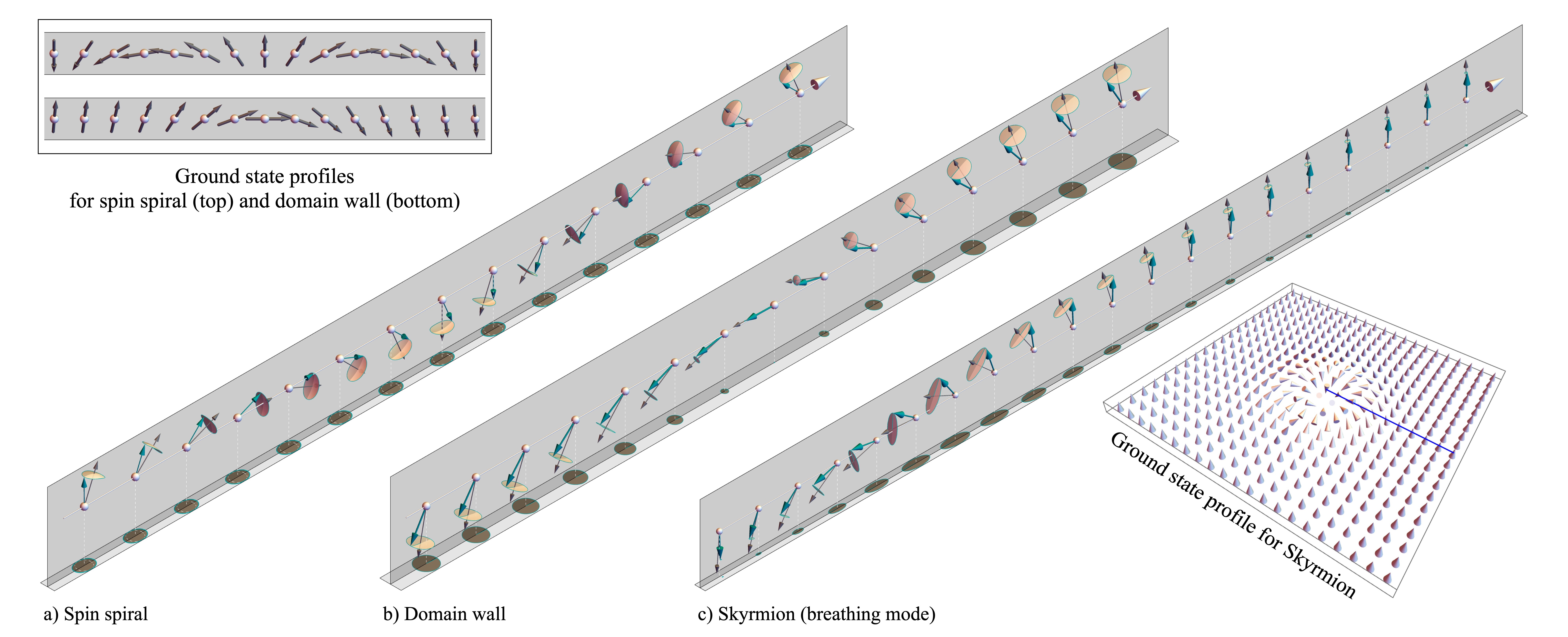}
    \caption{Simulated spin wave in: (a) spin spiral, (b) ferromagnetic domain wall, (c) N\'{e}el and Bloch type ferromagnetic Skyrmion. Top left inset: the ground state profile for spin spiral in (a) and domain wall in (b). Lower right inset: the ground state profile for the Skyrmion in (c), the profile in (c) is along a radial line cut indicated by the blue line.}
    \label{fig:sim2}
\end{figure*}

\emph{Dzyaloshinskii-Moriya interaction (DMI) - }
The spin spiral is a typical non-collinear magnetic texture which is stabilized by the competition between the Heisenberg exchange interaction and the Dzyaloshinskii-Moriya interaction (DMI). We consider the spin wave excitation of a spin spiral on a 1D magnetic chain with free energy
% \jx{I think we can neglect the $K$ term. Because the purpose of $K$ here is to kill the rotational Goldstone mode. But even with $K$, there is another translational Goldstone mode. We have to eliminate $K$ for the discussion of squeezing, because $K$ also contribute to squeezing. }
% \wyt{Ok. I remove the $1/2$ below to keep the consistency with the previous analysis.}
\begin{equation}
    F = \sum_{n=1}^N %\frac{K}{2}(\mb_n\cdot \hbz)^2 
    -D\hbz\cdot(\mb_n\times \mb_{n+1})-J\mb_n\cdot \mb_{n+1},
\end{equation}
%where $\hbz$ is a hard axis to ensure the magnetization lying in the $x$-$y$ plane, and 
where the DM vector is in $\hbz$ so that the magnetization spirals in the $x$-$y$ plane. 
% \jx{Shall the DM in $\hby$ and the spirals be in the $x$-$z$ plane?}
% \wyt{There will be no difference in this case. It is equivalent to rotate the coordinate system.}
The pitch of the spiral is determined by the relative strength of the DMI and Heisenberg exchange. The ground state with $\mb_n^0 = (\cos n\eta,\sin n\eta,0)$ and $\tan\eta = D/J$ is schematically depicted in the top left inset in \Figure{fig:sim2}.

Both Heisenberg exchange and DMI play roles in determining the polarization of spin wave excitation in a spin spiral. But their effects are the opposite: the Heisenberg exchange tends to polarize the spin wave perpendicular to the texture plane according to Rule \#2 (just as in the ferromagnetic ring case in \Figure{fig:sim1}(b)), while the DMI tends to polarize the spin wave within the texture plane according to Rule \#3.a. Overall, the effect of DMI surpasses that of Heisenberg because 
\begin{equation}
    \label{eqn:s_ss}
    s = s_\text{DM} + s_\text{H} 
    = D \sin\eta - 2J\sin^2\frac{\eta}{2}
%    = \frac{D}{2}\tan\frac{\phi}{2} 
    = \sqrt{D^2+J^2}-J > 0.
\end{equation}
% And $\epsilon = D\sin\eta + J \cos\eta$.
The simulated profile for the low wavelength (lowest frequency) spin wave mode in spin spiral is shown in \Figure{fig:sim2}(a), where we find that the spin wave is polarized within the texture plane, as expected by the competition  between Rule \#2 and \#3.a in \Eq{eqn:s_ss}.
% \jx{This is interfacial type DMI, what about the bulk type?}

% For the long wave length limit ($q \ra 0$),
% Similar to the macrospin case above, the overall positiveness of $s_0$ means that the spin wave tends to elongate in the $x$-$y$ plane.
% \jx{Rewrite the discussion about the polarization from $s_0$, and compare with qualitative argument in section II.}
% In this case, since both DMI and Heisenberg play roles simultaneously. The Heisenberg exchange still squeezes the spin wave out of the texture plane. However, the DMI tends to squeeze the spin wave into the texture plane, which coincides with the plane favored by the DMI vector. This is always the general tendency that DMI squeeze the polarization into a plane favored by the DM vector, not necessary to be the texture plane.

\emph{Circular polarization in Ferromagnetic Domain Wall - }
We now consider the most widely discussed magnetic texture -- the magnetic domain wall, for which we consider a 1D ferromagnetic chain with free energy 
\begin{equation}
   F = \int dx \qty[
   - \frac{K}{2}(\mb\cdot \hbz)^2 
%    \rmv{+ \frac{K'}{2}(\mb\cdot \hby)^2} 
   + \frac{A}{2}(\nabla \mb)^2].
\end{equation}
Here $K$ is the easy-axis anisotropy along $\hbz$, $A > 0$ is the ferromagnetic exchange coupling. A N\'{e}el type domain wall with equilibrium magnetization $\mb_0(x) = \bn_{\theta,0}$ is stabilized with the Walker profile $\theta = 2\arctan[\exp(x/w)]$ and domain wall width $w = \sqrt{A/K}$ \cite{schryerMotion180degDomain1974}. 
% \jx{Cite this reference here: Schryer, N. L. \& Walker, L. R., Journal of Applied Physics 45, 5406 (1974). }
The profile for the domain wall ground state is schematically depicted in the top left inset in \Figure{fig:sim2}. 
In this case, both uniaxial anisotropy and Heisenberg exchange play roles in determining the polarization of spin wave excitations. Initially, one might anticipate a non-circular spin wave polarization due to the non-trivial magnetic texture. However, the polarizing effects of anisotropy, which favors polarization within the texture plane (Rule \#1), and Heisenberg exchange, which promotes polarization perpendicular to the texture plane (Rule \#2), oppose one another. Consequently, the competition between the two determines the overall polarization of the spin wave across a domain wall. Remarkably, as seen in the simulated result in \Figure{fig:sim2}(b) for the low wave-length mode, the spin wave excitation exhibits a perfect circular polarization throughout the domain wall, suggesting that the polarizing influences of anisotropy and exchange effectively cancel out exactly.

To prove the exact cancellation, let's calculate the squeezing parameter due to the anisotropy and exchange separately. Since the angle between the local equilibrium magnetization $\mb_0(x)$ and the anisotropy axis $\hbz$ is $\eta = \theta(x)$, the squeezing parameter at position $x$ due to anisotropy is given by \Eq{eqn:sK}:
\begin{equation}
    \label{eqn:tani}
    s_\text{K}(x) 
    = -\frac{K}{2} \sin^2\eta = -\frac{1}{2}K\sin^2\theta(x).
\end{equation}
The squeezing parameter due to Heisenberg exchange given by \Eq{eqn:sH} can be obtained by considering the texture curvature at position $x$ using the continuous limit: with relative angle between two local equilibrium magnetization $\eta = a\partial_x \theta(x)$ and $J = A/a^2$ with $a$ the lattice constant, therefore 
% \wyt{I remove the $1/2$ here, then the coefficient is correct in the following equation.}
\begin{equation}
    \label{eqn:ttex}
    s_\text{H}(x)
    = 2J\sin^2\frac{\eta}{2}
    = \frac{1}{2}A\qty[\partial_x\theta(x)]^2.
\end{equation}
The squeezing parameters due to anisotropy \Eq{eqn:tani} and Heisenberg exchange \Eq{eqn:ttex} have opposite signs, indicating their competition in determining the spin wave polarization. Furthermore, the stability of magnetic domain wall with Walker profile is defined by identity $K\sin^2\theta(x)=A\qty[\partial_x \theta(x)]^2$ \cite{schryerMotion180degDomain1974},
% \jx{Use this reference here instead of [3]: Schryer, N. L. \& Walker, L. R., Journal of Applied Physics 45, 5406 (1974). }
which means the cancellation between $s_\text{K}$ \Eq{eqn:tani} and $s_\text{H}$ \Eq{eqn:ttex} is exact. Consequently, the squeezing of polarization vanishes everywhere, allowing the spin wave to maintain its circular polarization across the magnetic domain wall, irrespective of any deviations from the easy axis or the presence of inhomogeneous textures.

\emph{Ferromagnetic Skyrmion - }
A more complex magnetic texture is the magnetic Skyrmion in a 2D ferromagnetic thin film with free energy 
% \jx{Check the free energy expression.}
% \wyt{I add a $1/2$ on the DMI term, so the effective field is $D\nabla\times \mb$. This this bulk type DMI, and the Skyrmion is then Bloch type.}
\begin{align}
    F = \int d^2\br 
        &\Big\{- B \mb\cdot\hbz
        - \frac{K}{2}(\mb\cdot \hbz)^2 + \frac{A}{2}(\nabla \mb)^2 \nn
        &-D \qty[\mb\cdot\nabla(\hbz\cdot \mb)-(\nabla\cdot \mb)(\hbz\cdot\mb)]\Big\},
        % + \frac{D}{2} \mb\cdot(\nabla\times\mb)\right]
\end{align}
where $K$ and $A$ are the easy-axis anisotropy along $\hbz$ and the exchange coupling constant, $D$ is strength for the (bulk type) Dzyaloshinskii-Moriya interaction. The Skyrmion ground state as depicted in the bottom right inset in \Figure{fig:sim2} is a N\'{e}el type Skyrmion \cite{wangFundamentalPhysicsApplications2022}.
% \wyt{The Skyrmion of the free energy (20) is a Bloch type Skyrmion.}
% \jx{It is easier to visualize a Neel Skyrmion. Can we change free energy for a Neel type.}
Skyrmions are capable of accommodating confined soft modes, such as the breathing and twisting modes \cite{mruczkiewiczAzimuthalSpinwaveExcitations2018}, whose polarization relies on all types interactions in this system, \ie anisotropy, Heisenberg exchange, and Dzyaloshinskii-Moriya interaction (DMI). The anisotropy tends to align spin waves predominantly in the plane spanned by $\hbz$ and the radial direction (Rule \#1). When focusing solely on the radial texture and neglecting angular variations, DMI further supports this radial polarization (Rule \#3.a). However, the Heisenberg exchange typically promotes the polarization perpendicular to this plane (Rule \#2). Therefore, the overall polarization relies on the competition of the anisotropy and DMI against the Heisenberg exchange.  Nevertheless, the predominance of anisotropic effects combined with DMI leads to a net polarization that aligns within the radial texture plane, overshadowing the influence of Heisenberg exchange. \Figure{fig:sim2}(c) shows the numerical simulation for the breathing mode of a Skyrmion based on COMSOL Multiphysics \cite{zhangFrequencydomainMicromagneticSimulation2023}, where we find the spin wave is polarized in the radial direction.

\section{Conclusions}

In conclusion, we have established a set of rules that govern the polarization of spin waves in ferromagnetic materials based on the competition between the anisotropy, Heisenberg exchange, and Dzyaloshinskii-Moriya interaction. These rules provide a semi-quantitative understanding of the spin wave polarization in diverse magnetic textures. We have verified these rules through numerical simulations of spin wave excitations in different magnetic textures, and found that the spin wave polarization is consistent with the established rules. Our findings provide a comprehensive understanding of the spin wave polarization in magnetic textures, which is crucial for the design and manipulation of spin wave excitations in magnetic devices.
% \jx{Let's just keep a simple conclusion, and omit the discussion. This is an AI generated conclusion paragraph, please polish the conclusion paragraph.}
% \wyt{I think the expression of this paragraph is good enough if we want to make it simple. I also ask GPT, kimi and deepseek whether it is like an AI generated one. They all answer no.}
% An interesting aspect of the spin wave profile across a domain wall is the phenomenon where the amplitude of the spin wave diminishes significantly at the center of the domain wall (see \Figure{fig:sim2}(b)). This observation raises questions regarding the mechanism of energy transmission across the wall despite the near absence of excitation at this point. The explanation lies in the nature of magnetic excitation energy; it is not predominantly derived from the misalignment between the excited magnetization and the local equilibrium magnetization. Instead, part of the magnetic energy is stored in the exchange interactions, enabling efficient energy transfer even when spin wave amplitude appears minimal at the domain wall center.
% \jx{This seems to indicate that exciting spin wave at the domain wall center is not very efficient. }
% \add{Anything else to discuss? Non-linear regime?}

\bigskip

{\it Acknowledgements.} 
This work was supported by 
National Natural Science Foundation of China (Grants No. 12474110),
the National Key Research and Development Program of China (Grant No. 2022YFA1403300),
the Innovation Program for Quantum Science and Technology (Grant No.2024ZD0300103),
and Shanghai Municipal Science and Technology Major Project (Grant No.2019SHZDZX01).

\bibliography{ref,spinwave}
\end{document}